\documentclass{article}

\usepackage{arxiv}

\usepackage[utf8]{inputenc} 
\usepackage[T1]{fontenc}    
\usepackage{hyperref}       
\usepackage{url}            
\usepackage{booktabs}       
\usepackage{amsfonts}       
\usepackage{nicefrac}       
\usepackage{microtype}      
\usepackage{lipsum}
\usepackage{amssymb, bm, amsmath}
\usepackage{mathtools}

\newcommand{\quotes}[1]{``#1''}

\title{Generosity, selfishness and exploitation as optimal greedy strategies for resource sharing}

\author{
  Andrea Mazzolini \\
  Quantitative life science Unit\\
  The Abdus Salam International Center for Theoretical Physics\\
  Trieste, Italy \\
  \texttt{amazzoli@ictp.it} \\
   \And
  Antonio Celani \\
  Quantitative life science Unit\\
  The Abdus Salam International Center for Theoretical Physics\\
  Trieste, Italy \\
  \texttt{acelani@ictp.it} \\
}

\begin{document}
\maketitle

\begin{abstract}
Resource sharing outside the kinship bonds is rare.
Besides humans, it occurs in chimpanzee, wild dogs and hyenas, as well as in vampire bats.
Resource sharing is an instance of animal cooperation, where an animal gives away part of the resources that it owns for the benefit of a recipient.
Taking inspiration from blood-sharing in vampire bats, here we show the emergence of generosity in a Markov game, which couples the resource sharing between two players with the gathering task of that resource.
At variance with the classical evolutionary models for cooperation, the optimal strategies of this game can be potentially learned by animals during their life-time.
The players act greedily, that is, they try to individually maximize only their personal income. 
Nonetheless, the analytical solution of the model shows that three non trivial optimal behaviours emerge depending on conditions. 
Besides the obvious case when players are selfish in their choice of resource division, there are conditions under which both players are generous. 
Moreover, we also found a range of situations in which one selfish player exploits another generous individual, for the satisfaction of both players.
Our results show that resource sharing is favoured by three factors: a long time horizon over which the players try to optimize their own game, the similarity among players in their ability of performing the resource-gathering task, as well as by the availability of resources in the environment. These concurrent requirements lead to identify necessary conditions for the emergence of generosity.
\end{abstract}

\section{Introduction}

%
%
The sharing of resources in animals can be defined as an active or passive transfer of food from one individual to another \cite{stevens2004conceptual, jaeggi2011evolution, jaeggi2013natural}.
Some examples are food division in primates \cite{brown1978food, feistner1989food, boesch1989hunting, de1989food, mitani2006reciprocal}, blood sharing in vampire bats \cite{wilkinson1984reciprocal, wilkinson1990food, wilkinson2016non}, and cooperative breeding in birds and fishes \cite{koenig2004ecology, wong2011evolution}.
From an external human observer, this behavior really resembles an act of \quotes{generosity}.
Indeed, once an individual has acquired some resources, it chooses to equalize them among the members of the community (or with a partner) instead of being \quotes{selfish} and having a larger income.
%
%
Resource division can be viewed as an instance of animal cooperation and biological altruism, subjects that have a long tradition in evolutionary biology \cite{dugatkin1997cooperation, kappeler2006cooperation, okasa2013biological, barta2016individual}.
Darwin himself already noticed that altruism seems to be in contrast with the natural selection theory, since the altruistic act consists in a decrease of the donor fitness \cite{darwin1859origin}.
Many episodes of food sharing and altruistic behaviors occur between relatives, such as parent-offspring sharing \cite{feistner1989food}, and can be explained by kin-selection \cite{hamilton1964genetical, taylor1996make, foster2006kin}.
However, several other instances do not involve kin-related individuals, and require alternative explanations.
Some classical proposals are reciprocity \cite{trivers1971evolution}, repression of competition \cite{frank2003repression}, or ideas based on group selection \cite{wilson1994reintroducing}.

%
%
A concept tightly related to resource sharing is inequity aversion, defined as the tendency to negatively react in response to unequal subdivision of resources between individuals \cite{brosnan2014evolution, oberliessen2019social}.
One widespread behavior is the protest of the individual getting less than the partner, called first-order inequity aversion.
Animals that show this behavior are, for example, monkeys \cite{brosnan2003monkeys, takimoto2010capuchin}, dogs \cite{range2009absence} and crows \cite{wascher2013behavioral}.
In fewer cases, it is the individual receiving more than its fair share that protests, possibly acting in a way to equalize the resources.
This is dubbed second-order inequity aversion.
It seems that only chimpanzees show experimental evidence of this apparent \quotes{generous} behavior, by protesting if the partner receives less \cite{brosnan2010mechanisms}, or by choosing a fair resource division \cite{proctor2013chimpanzees}.
Importantly, inequity-averse behaviours are all inefficient for the short-term income: the individual that protests typically lowers its resource gain.
A widely used theoretical framework for describing inequity-aversion is the ultimatum game \cite{guth1982experimental, camerer2003behavioral}.
It is played by a \textit{proposer} who has to divide a certain amount of resource with a \textit{receiver}.
This second player has the possibility to accept the proposal or to reject it, in the latter case neither party receives anything. 
Note that a fair choice of the proposer (i.e. dividing the resource in half) corresponds to the $2^{nd}$ order inequity-averse behavior, while the rejection of an unfair proposal to the $1^{st}$ order one.
A lot of theoretical work has been done to understand the emergence of fairness in the ultimatum game \cite{debove2016models}, since a one-shot game would always lead to the receiver's acceptance of the division with the minimum possible amount of resource for the receiver \footnote{This is the only evolutionary stable strategy.}.
Some mechanisms that can lead to the emergence of fairness are based on the knowledge of the players' reputation \cite{nowak2000fairness, andre2011social}, noise in the decision rule (with a larger noise for the receiver) \cite{gale1995learning}, spite \cite{huck1999indirect, forber2014evolution}, and spatial population structure \cite{page2000spatial, alexander2007structural, iranzo2011spatial}.
%
%
Most of these models as well as the ones for the emergence of cooperation are based on the evolutionary point of view. They address the question on how natural selection has led to a successful \quotes{cooperative/fair phenotype}, i.e. having a higher fitness than the one of defectors.
The natural framework to describe this kind of processes is evolutionary game theory, where one looks for the conditions in which the cooperation is stable against the emergence of defectors \cite{axelrod1981evolution, nowak2006five, debove2016models}.

The present work is based on a different approach: we assume that animals have the ability to learn and change their strategies in the most efficient way for their survival in a given environment.
This allows us to move the focus from the time-scale of evolution of a large population to the scale of an individual life span, who interacts with relatively few other individuals within its community.

In particular, we adopt Markov decision processes, \cite{sutton1998introduction}, which, in the case of more than one agent, are called Markov games \cite{littman1994markov, hu1998multiagent, claus1998dynamics}.
%
%
In short, in this framework agents know the environmental state in which they are, and, depending on this state, can choose actions according to a policy.
As a consequence of these actions, the agents receive rewards (here represented by resources), and the game moves to a new environmental state which depends on the player actions and the previous state.
Each agent wants to find the policy that maximizes its resource income within a given time horizon.

%
%
There are few works that study non-kin cooperation using reinforcement learning instead of evolution \cite{sandholm1996multiagent, claus1998dynamics, ezaki2016reinforcement}.
They focus on the emergence of optimal strategies mainly in the context of the iterated prisoner dilemma.
Differently, the Markov model introduced in the following takes inspiration from resource-gathering and sharing tasks between two animals, and, in particular, from the blood sharing in vampire bats, as described in the next section.
Generous resource sharing, or a $2^{nd}$ order inequity-aversion behavior, emerges if the optimal policy for the player is to divide a certain amount of acquired resources in equal parts instead of keeping a larger fraction.
The model allows for analytical solution and, albeit its simplicity, exhibits a rich phenomenology of the optimal strategy, i.e. being generous or selfish, depending on few parameters.
These key parameters have a straightforward biological interpretation: the abundance of resource in the environment, the player asymmetry in the ability of performing the tasks (typically overlooked in the traditional literature of cooperation), the number of games that a player expects to play, i.e. the time horizon, and the fraction of resources kept in the selfish choice.
Specifically, three different phases, corresponding to generous, exploitative and selfish behaviour, can be discerned.
%
%
It is important to stress that individuals are \quotes{greedy}, meaning that they are interested in obtaining the largest possible personal gain, without considering the resource income of their partners.
This clearly implies that, if one were looking at the immediate reward only, it would be always optimal to be selfish.
However, since individuals are interested in the return within a certain time horizon, generosity can provide advantages in the future, becoming, possibly, the most efficient strategy even for \quotes{greedy} players.

\section{A relevant example: food sharing in vampire bats}

\begin{figure}
\includegraphics[width=0.6\textwidth]{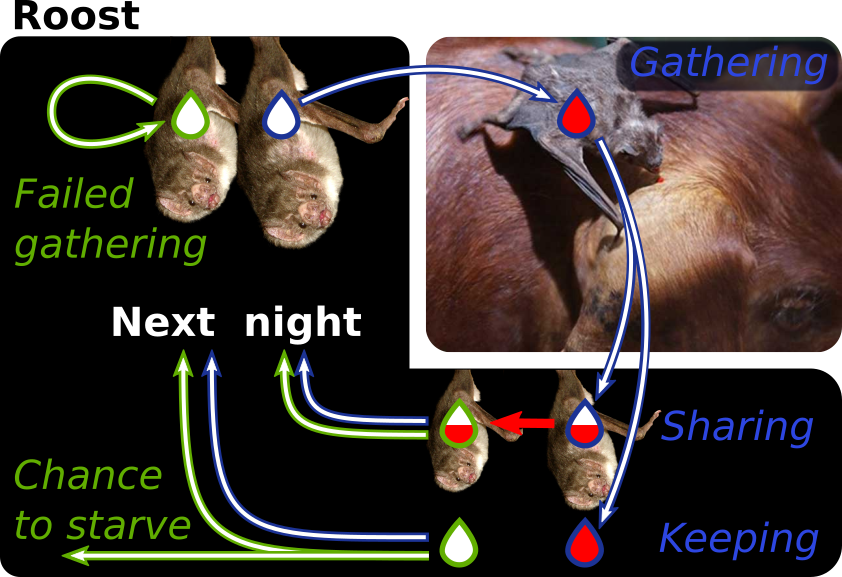}
\centering
\caption{Schematic illustration of the blood gathering and sharing shown by vampire bat communities.}
\label{fig:bats}
\end{figure}

The model that we will introduce is mainly inspired by the well-studied \quotes{generous} behaviour shown by blood-feeding vampire bats \cite{wilkinson1984reciprocal, wilkinson1990food, wilkinson2016non}.
After gathering blood during a given night, these animals frequently share it to other individuals when they come back to the roost. 
This behaviour is shown most of the times by mothers feeding their own offspring, and can be explained by kin selection. 
However a relevant fraction of blood sharing occurs among unrelated adults.
The key factors that seems to be at the basis of this behaviour are, first, the fact that after two consecutive nights without blood a bat can reach starvation, and, second, that there is around $18\%$ of probability that an individual fails to obtain blood during the night  \cite{wilkinson1984reciprocal}.
In the light of this, food sharing seems to be necessary to prevent starvation, and, on the long run, the collapse of the community.
Therefore, even though the individual fitness decreases in the short-term period, being generous may provide a significant long-term advantage, since it guarantees the survival of the community.

Inspired by these findings, in this paper we will build a minimal model of resource sharing that retains the fundamental ingredients leading to this behaviour.
The game is between two individuals that can interact repeatedly in time. This requires that the animals are able to discriminate between different individuals within the community, since, in principle, different strategies can be learned depending on the partner. 
This seems to be true in bats: specific pairs can establish long-term relationships by helping each other \cite{wilkinson1990food}.
It is important to stress that the only information that the bat needs about the parter is its health state. They do not need to remember, for example, its actions at the previous stage, as a Tit-for-Tat strategy would require \cite{axelrod1981evolution}, which is the classical mechanism through which reciprocal cooperation is obtained.

One stage of the game corresponds to one night, and describes the situation in which one individual gathers some blood, and the second fails in performing this task or stays in the roost, as shown in the upper part of Figure \ref{fig:bats}.
At subsequent stages of the game the two roles can be inverted, and this depends on a given probability that can account for a possible \quotes{asymmetry} between the players: one can be the gatherer more frequently than the other one.
Actually, this is also observed in field studies regarding chimpanzees, where, for example, the degree of dominance within the community, the age, and the sex determine the frequency at which animals perform resource-gathering tasks \cite{boesch1989hunting}.
After having obtained some food, the gatherer comes back to the roost and chooses whether to share it with individuals that have failed the task, or to keep most of it for itself (bottom right part of Figure \ref{fig:bats}).
As a result of this division, the two individuals feed themselves, and their health changes accordingly to the amount of food received (bottom left part of Figure \ref{fig:bats}). 
As in real bats, if the individual that failed the gathering task does not receive enough food, it may starve and, as a consequence, it will not be able to gather resources anymore.

\section{A minimal decision-making model for resource gathering and sharing}
\label{sec:model}

\begin{figure}
\includegraphics[width=0.6\textwidth]{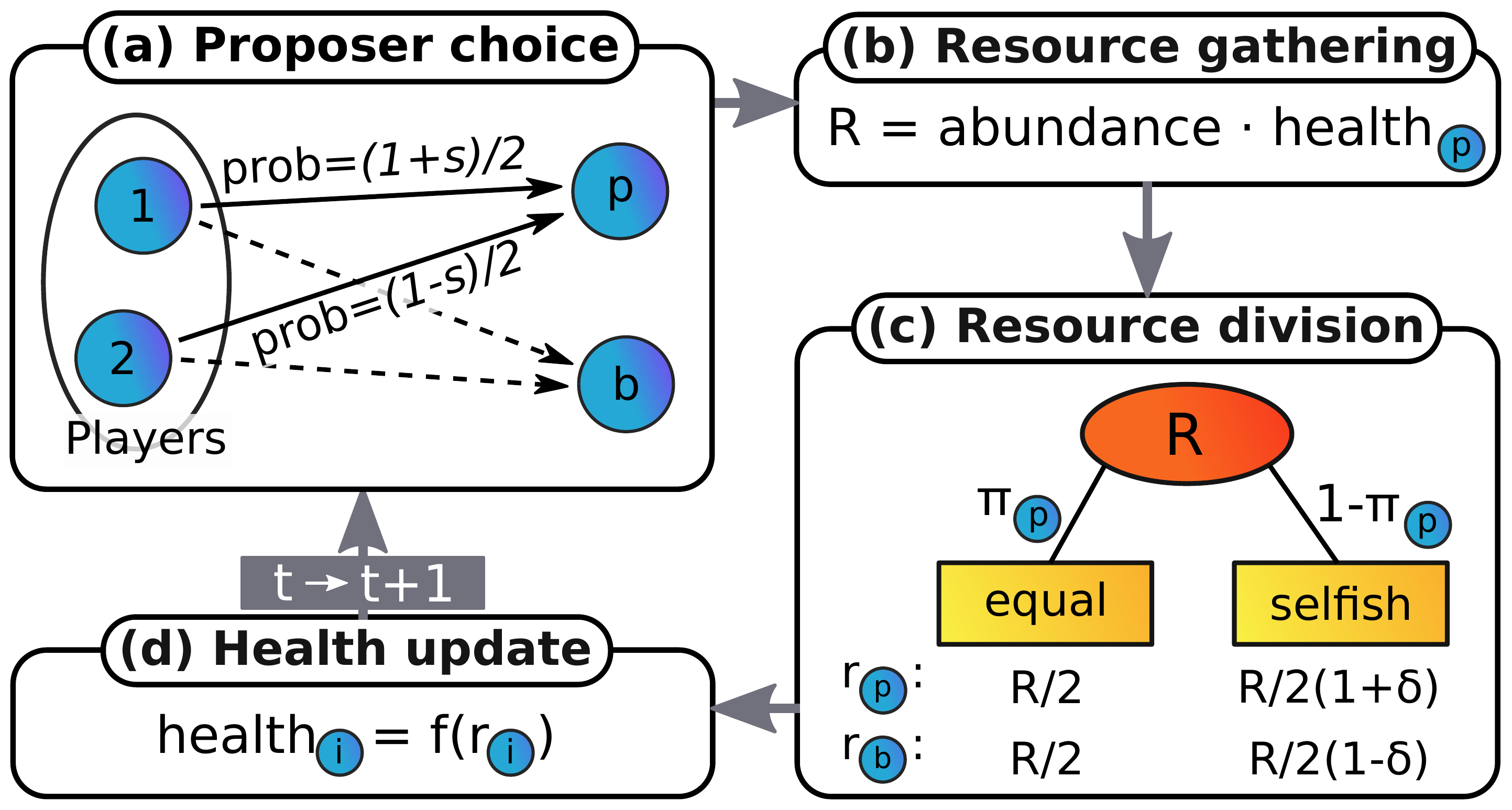}
\centering
\caption{Illustration of one temporal step of the resource-gathering-sharing model. Box (a): one of the two players is chosen to be the \textit{proposer}, labelled with $p$, with a certain probability depending on the first-player specialization $s \in \left[ -1, 1 \right]$.
The other individual becomes the \textit{bystander}, $b$. 
Box (b): the proposer performs the resource gathering task, acquiring a given amount of resource $R$ depending on its internal variable $h_p$, which can be thought as its health, and the resource abundance $a$.
Box (c): playing a dictator game, the proposer chooses how to divide the resource with the bystander.
It can either split $R$ in equal parts, $r_p = r_b = R/2$, or be selfish and keep a larger fraction of resource, $r_p=R/2(1+\delta)$, where $\delta \in (0,1]$. 
The proposer strategy is determined by its policy $\pi_p$, i.e. the probability of playing the generous action and split the resource in equal parts.
Box (d): finally, each player updates its health, $h_i$, as a function of the received resource fraction $r_i$.}
\label{fig:model}
\end{figure}

Here we describe the model in abstract mathematical terms. 
Even though it is inspired by the behaviour of vampire bats, it can be used as a backbone in whatever context where  generous resource sharing emerges.
%
It is performed by two animals/players, whose only purpose is to acquire as many resources as possible for themselves. 
At each iteration of the game (which we are going to call also \quotes{episode}), one of the two players is selected to be the \textit{proposer}, $p$, box (a) of Figure \ref{fig:model}: it will be the active individual who carries out a certain task to acquire resources.
The other player, called the \textit{bystander}, $b$, remains passive until the end of the iteration.
The model features a degree of specialization of the players, such that at every iteration, one individual can be consistently more likely to play as proposer than the other individual.
Specifically, the first player can be selected with probability $(1+s)/2$.
The parameter $s \in [-1,1]$ can be interpreted as the \textit{specialization} of the first player with respect to the second one for the given task: $s=1$ means that the first player will always perform the task, while, $s=0$ defines a symmetric case where both can be chosen with the same probability.
%
After being selected, the proposer performs the task, Figure \ref{fig:model}b, which consists in acquiring a certain amount of resource $R = a \cdot h_p$, depending on the abundance of the resource in the environment $a$, and an internal variable of the proposer $h_p$.
The variable $h$ can be identified with the player health.
In the interest of simplicity, we consider $h$ to be a binary variable: $h \in \lbrace 0,1 \rbrace$.
The value $h=0$ defines an exhausted agent, who will not gather any resource, while $h=1$ a healthy one.
%
Typically, individuals that get food are also the ones that have more decision power in the resource sharing \cite{wilkinson1984reciprocal, boesch1989hunting}.
Therefore we let the same player that has performed the task also to choose how to share the resource with the bystander.
\\

The resource-division stage, Figure \ref{fig:model}c, consists in a mini-dictator game \cite{camerer2003behavioral}, a simplified version of the ultimatum game \cite{guth1982experimental}.
It is worth pointing out that using the more complex ultimatum game would not change the optimal strategies, as shown in \ref{sec:UG}.
Specifically, the game consists in two actions for the proposer: splitting the resource $R$ in equal parts, i.e. the \quotes{generous} action, such that he obtains $r_p = R/2$, or being \quotes{selfish} by keeping a larger fraction $r_p = R/2(1+\delta)$, with $\delta \in (0,1]$.
The bystander is passive and will receive resources according to the proposer choice: $r_b = R/2$ for the equal action, and $r_b = R/2(1-\delta)$ for the unequal.
The exogenous parameter $\delta \in (0,1]$, called \textit{inequity strength}, indicates how large is the selfish-choice profit for the proposer at the expense of the bystander.
It can be interpreted as a constraint on the maximum amount of resource that a player can keep or control. 
For example, the case $\delta=1$ defines a game in which the proposer is able to consume all of the resource or, alternatively, it is able to protect it from the attempted theft of the bystander.
The decision of a player, i.e. being \quotes{generous} or \quotes{selfish}, is encoded by the policy $\pi$, which defines the probability of playing the equal action.
This is the parameter that each individual can control in order to maximize the amount of obtained resource.
%
The final step consists in updating the health according to the received fraction of resource $r$, Figure \ref{fig:model}d.
One simple choice to express the health dependency on $r$ is the Heaviside step function $f(r) = H(r-\theta)$, which is $1$ if $r \ge \theta$, or $0$ if $r < \theta$.
In order to be healthy, the player has to receive an amount of resource larger than the threshold $\theta$, which represents the minimal food ration that a player needs to be in good shape.

\begin{table}
\centering
\caption{Key parameters of the model.}
\begin{tabular}{rlrl}
$s$ & First-player specialization & $\delta$ & Inequity strength \\
$a$ & Resource abundance & $\theta$ & Food threshold \\
$\gamma$ & Discount factor & $(1-\gamma)^{-1}$ & Time horizon \\
\end{tabular}
\label{tab:params}
\end{table}

%
The objective function to be maximized by player $i$ is the exponentially discounted return:
\begin{equation}
G_i (\pi_i, \pi_{-i}) = \mathbb{E}_{\pi_i, \pi_{-i}} \left[ \sum_{t=1}^\infty r_i^{(t)} \gamma^{t-1} \right] ,
\label{eq:return}
\end{equation}
which is averaged over the stochastic game evolution, the player policy $\pi_i$, and the other player policy $\pi_{-i}$.
The random variable $r_i^{(t)}$ indicates the fraction of resource obtained at the iteration $t$ of the model, and $\gamma \in [0,1)$, the so-called \textit{discount factor}, determines the value of future rewards, such that increasing $\gamma$, the player gives more importance to what it is going to earn at future time steps.
Since $1 - \gamma$ can be interpreted as the probability that the game ends at each iteration, it sets the average length of a game to $(1-\gamma)^{-1}$, the so-called \textit{time horizon}, defining the number of episodes over which players want to maximize the return.
Each individual can control its own policy $\pi_i$ affecting the outcome of resource fraction $r_i^{(t)}$, both for himself and the other individual.
Here, solving the problem for the player $i$ means finding the policy $\pi_i$ which maximizes its own profit $G_i$ by knowing that the other player $j$ is trying to optimize at the same time $G_{j}$.
The obtained optimal policies belong to a Nash equilibrium: any unilateral change of strategy by a single player would reduce the payoff.
Here we also want to stress that the Markov property of the model implies that players take decisions only on the basis of the present state, and independently on previous decisions or game history.
This problem can be formalized using Bellman optimality equation as shown in the Methods.
It turns out that the model admits an analytical solution, and we were able to find the optimal strategies $\pi^*_i$ depending on the parameters of the model, as shown below.

%
%
Before presenting the solution, we would like to point out the general guiding lines that have driven the model construction.
The first important observation is that the optimal strategy of a simple dictator game is always to play selfishly (see \ref{sec:generalization}).
Therefore, we had to introduce an additional reasonable mechanism that allows generosity to emerge.
By taking inspiration from animal behavior, we paired the resource sharing with the gathering of that resource, using the health as the variable that couples the two tasks.
Specifically, the gathering efficiency depends on the health, which, in turn, is affected by how the sharing of the resources is performed.
This coupling is what we assumed to be the basic mechanism to obtain equal resource sharing.
Starting from this idea, we aimed at building a Markov game with a \quotes{minimal mathematical complexity}.
Therefore, we chose the most simple possible sharing task, i.e. a dictator game, a gathering task where the obtained resource is proportional to the health, and a health which is a binary variable updated with a step function.
In general, the purpose of keeping the complexity low is to allow mathematical techniques to provide more insights about the model solution.
In our case, the chosen structure is simple enough to explicitly derive an analytical solution.
The fact that there are four free parameters seems to go against the purpose of keeping the complexity low, however, since the model remains solvable, one can have a full control on how the solution depends on them.
Moreover, since all of them have a reasonable biological interpretation, the dependence of the solution on those parameters allows one to identify general rules that govern the generous-selfish dilemma, as discussed later.

\section{Results}
\subsection{Inequity aversion is a matter of time horizon and player specialization}

\begin{figure}[t]
\centering
\includegraphics[width=\textwidth]{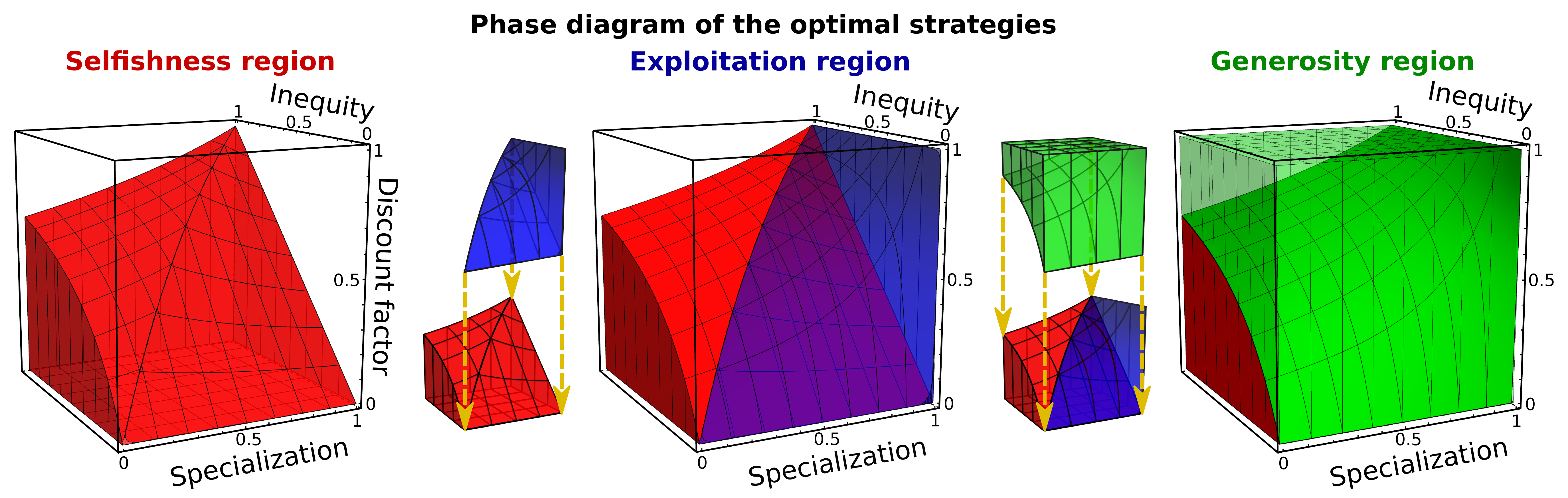}
\caption{Optimal strategies for the maximization of the exponentially discounted return (\ref{eq:return}) as a function of the three model parameters: the player specialization $s$, the discount factor $\gamma$, and the inequity strength of the selfish choice $\delta$.
Looking at the panel (a), starting from the left plot, the red volume is defined by Equation (\ref{eq:selfish_ph}), and represents the region of parameters such that the selfish strategy is optimal: $\pi_1^* = 0, \pi_2^* = 0$.
The blue volume shown in the central diagram, which rests on the selfish block, defines the \quotes{exploitation} phase: $\pi_1^* = 0, \pi_2^* = 1$, Equation (\ref{eq:expl_ph}).
Above these two regions, the optimal strategy is the generous one, $\pi_1^* = 1, \pi_2^* = 1$, represented by the green volume in the third plot, Equation (\ref{eq:generous_ph}).}
\label{fig:ph_diag}
\end{figure}

%
%
Even though each player wants to greedily maximize its own profit, the generous strategy, i.e. $\pi_i^* = 1$, turns out to be the most efficient one in a broad region of the parameter space.
To intuitively understand why, one must note that an equal resource division can result in a higher long-term reward for the proposer.
Indeed, the generous choice provides more food to the other player, who has a higher chance to be \quotes{healthy} at the end of the episode (this can or cannot happen depending on the model parameters).
As a consequence, if it will play as proposer at future episodes, it could gather a larger amount of resource which will be shared with the former proposer.
This benefit typically disappears for the selfish choice: the other player gets few resources becoming \quotes{exhausted}, which, in turn, implies that the amount of food that it could get at future gathering tasks will be zero.
Therefore the optimal strategy must be a balance between this long-term benefit that a generous strategy can provide, and the immediate higher profit given by a selfish division.

%
%
Before describing the optimal solution, it is important to note a useful symmetry of the model: the game is invariant by changing the sign of the player-specialization $s$ and exchanging the first with the second player.
This allows us to consider only half of the player-specialization domain $s \in [0,1]$: all the results at negative $s$ can be immediately recovered by taking the result for the positive $s$ and inverting the players.

%
%
Figure \ref{fig:ph_diag} shows which strategy maximizes the objective function (\ref{eq:return}) varying the three free parameters $s$, $\delta$, $\gamma$, and considering only half of the specialization interval for the reason described above.
The role of the resource abundance, $a$, and the minimal food threshold, $\theta$, are discussed in the next section, and here fixed at $a = 1$, and $\theta = 1/2$. 
Three distinct phases emerge as optimal strategies: \textit{selfishness}, where both players choose to keep a larger fraction of resource, \textit{exploitation}, where one player is selfish and the other generous, and the case of two individuals that split the resource in equal parts, \textit{generosity}.
As derived in \ref{sec:solution}, the range of parameters in which being selfish is optimal, $\pi_1^* = 0, \pi_2^* = 0$, is given by the following inequality:
\begin{equation}
\text{Selfishness: } \;\; \gamma < \frac{2 \delta}{1 + |s| + 2 \max{(0, \delta - |s|)}} ,
\label{eq:selfish_ph}
\end{equation}
which defines the red volume in the Figure \ref{fig:ph_diag}a.
As the intuition suggests, the selfish behaviour emerges when players give much more importance to the short-term profit, i.e. having a sufficiently small discount factor $\gamma$.
Consistently, in the extreme case of $\gamma = 0$, where each player considers only the immediate reward, the selfish behaviour is always optimal.
%
%
A different scenario appears in the following region of parameters, represented in the second plot of Figure \ref{fig:ph_diag}b in blue:
\begin{equation}
\text{Exploitation: } \;\; \frac{2 \delta}{1 + s} < \gamma < \frac{2 \delta}{1 - s + 2 \delta} .
\label{eq:expl_ph}
\end{equation}
Here what is optimal for the first player is being selfish, while for the second one is being generous, $\pi_1^* = 0, \pi_2^* = 1$.
The two inequalities above also imply that $s > \delta$ (by verifying that the left-hand term is less than the right-hand one).
Therefore, if there is a strong asymmetry between the individuals, i.e. the first is much more specialized, playing as proposer with probability $p > (1+s)/2$, a dominance relationship of one player over the other can emerge.
Taking advantage of the model symmetry, and substituting $s \rightarrow -s$ in Equation (\ref{eq:expl_ph}), one can immediately recover the case in which the two roles are inverted: $\pi_1^* = 1, \pi_2^* = 0$, that exists for $s < -\delta$, i.e. a second player more specialized than the first one.
%
%
Finally, for a sufficiently large discount factor, the generous behaviour can be the more rewarding strategy, $\pi_1^* = 1, \pi_2^* = 1$.
Specifically, for:
\begin{equation}
\text{Generosity: } \;\; \gamma > \frac{2 \delta}{1 - |s| + 2 \delta} .
\label{eq:generous_ph}
\end{equation}
This region is shown in green in the third plot of Figure \ref{fig:ph_diag}c, which in the limit $\gamma \rightarrow 1$ becomes the optimal strategy for any choice of the other parameters (except $|s|=1$).

\begin{figure}[t]
\centering
\includegraphics[width=0.6\textwidth]{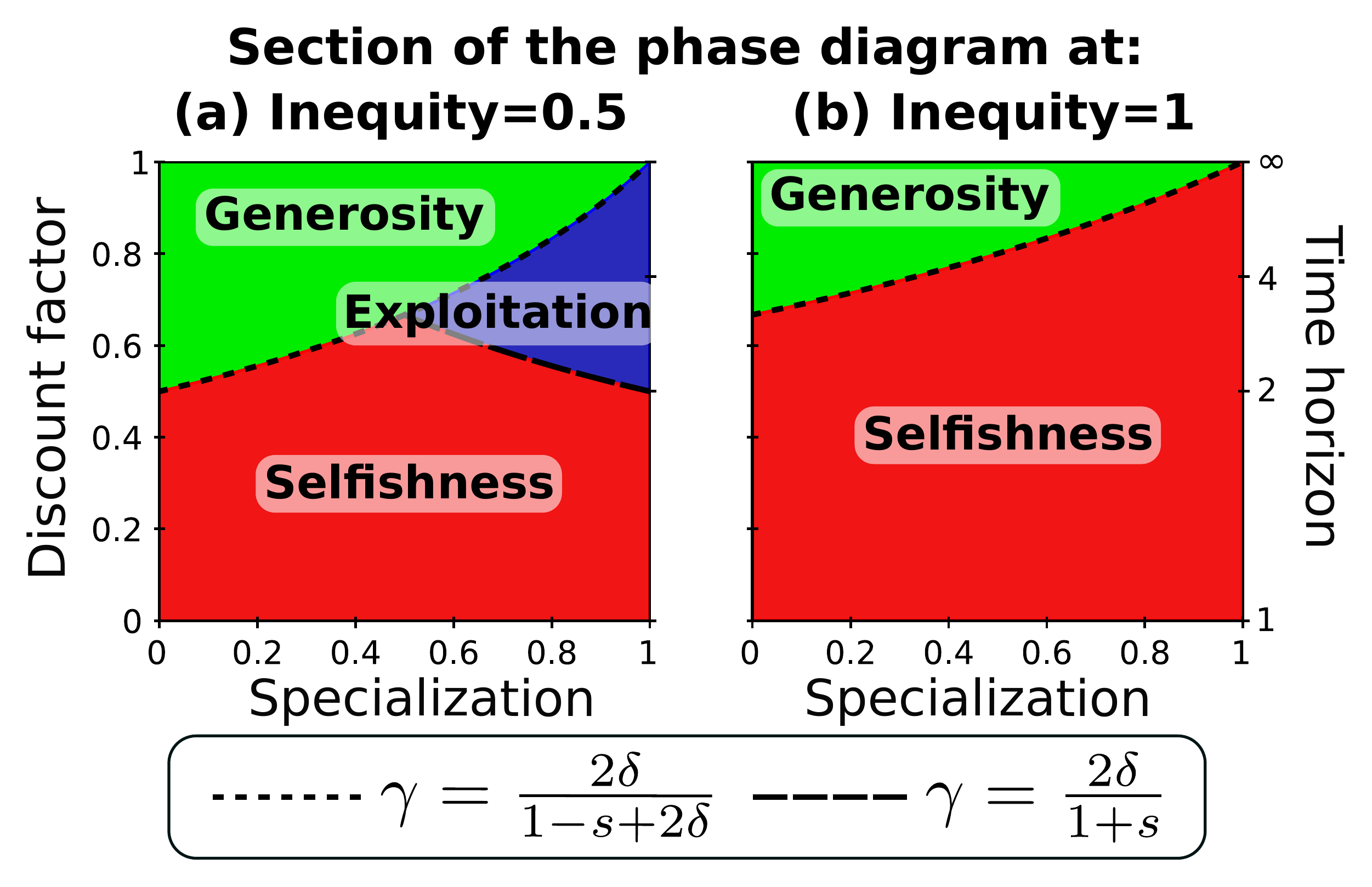}
\caption{Section of the three-dimensional phase diagram at two values of the inequity strength: $\delta = 1/2$ and $\delta = 1$.
The right axis refers to the time horizon $(1-\gamma)^{-1}$.}
\label{fig:ph_diag_section}
\end{figure}

The panel (a) of Figure \ref{fig:ph_diag_section} is useful to summarize the results discussed so far.
It represents a section of the three-dimensional phase diagram at fixed inequity strength $\delta=1/2$, therefore describing all the games in which the selfish resource division is $3/4 R$ for the proposer and $1/4 R$ for the bystander.
It is clear that the emergence of inequity aversion, i.e. the generous strategy, is favoured by a high discount factor, that is when players give more importance to the future rewards that a healthy mate can provide.
However, if the player specialization is sufficiently large, even for high $\gamma$, the game can enter the blue area.
In such a case, the first agent plays as proposer much more frequently than the second one, taking control of the game: it can be selfish because the potential future profit provided by the other player becomes low.
At the same time, the second player is going to be the bystander with very high probability, and therefore it has to play generously (in the unlikely case of being selected as proposer) in order to make the first one in a healthy state and receive at least $1/4$ at the next resource divisions.
Finally, when the discount factor goes below the line defined by Equation (\ref{eq:selfish_ph}), both the players start to be selfish, having an interest in maximizing the short-term profit.
Note that here the selfish strategy is always optimal if $\gamma < 1/2$, or, equivalently, the number of episodes that the players expect to play, i.e. the time horizon $(1-\gamma)^{-1}$, is less that $2$.
The limiting case of $\delta = 1$ is shown in panel (b) of Figure \ref{fig:ph_diag_section}.
Here the choice is between an equal division and keeping all the resource.
Since the gain of the selfish division increases with respect the previous case, the region associated with generosity reduces, but does not disappear.
Differently, exploitation does not longer represent an optimal behaviour, because the bystander of a selfish division does not receive anything, and therefore it is not interested in keeping the other player healthy.

\subsection{The role of resource abundance}

Environmental factors affect the amount of resources in the habitat, resulting in different outcomes of a gathering task, which are independent of the animal \quotes{health} or its ability to fulfil the task.
Moreover, changes of the available food amount can have strong effects on the sharing behavior, as shown, for instance, by chimpanzees in the Ta\"i National Park, which are more generous when the prey is large, typically an adult of a Colobus monkeys, with respect when they capture infants \cite{boesch1989hunting}.
In our minimal model, $a$ is the free parameter that can describe the quantity of resource provided by the environment.
The role that it plays in determining the optimal strategy is coupled with $\theta$, i.e. the minimal quantity of food that a player needs to be healthy (both the two parameters are kept fixed in the previous section, $a=1, \theta=1/2$).
Indeed, they determine how the health is updated: for example, high abundance and low food threshold should lead more easily to healthy players, and, in turn, this can change the optimal policies.
As shown in \ref{sec:ab}, the optimal solution is controlled by the ratio between the two parameters through the following inequalities:
\begin{equation}
\text{Necessary conditions for generosity: } \; \; 1 - \delta < \frac{2 \theta}{a} \le 1 .
\label{eq:abundance_ineq}
\end{equation}
If they are satisfied, the optimal behaviour is exactly described by the relations shown in the previous section: (\ref{eq:selfish_ph}), (\ref{eq:expl_ph}), (\ref{eq:generous_ph}). 
Otherwise, if one of the two inequalities is violated, the selfish behaviour ($\pi^*_i=0$) is the most efficient one for each parameter choice.

Intuitively, this happens because, if (\ref{eq:abundance_ineq}) is true, the generous resource division leads to a healthy partner (the health-update function, Figure \ref{fig:model}d, reads $H[a/2 - \theta] = 1$), while the selfish one drives him to exhaustion, $H[a/2(1-\delta) - \theta] = 0$.
This provides the long-term benefit for the equal division (and not for the unequal one) described at the beginning of the previous section, which is the key to have generosity.
On the contrary, if the environment provides too few resources such that (\ref{eq:abundance_ineq}) is broken, $a < 2 \theta$, then there is not enough food to make the other player healthy: the long term benefit disappears, implying that it is always convenient to keep as much food as possible through the unequal division.
Also in the case of a very rich environment, i.e. $a \ge 2 \theta / (1-\delta)$, the optimal policy is always being selfish.
In this case, there is enough food to make the bystander always healthy, even as a recipient of the selfish division.
As a consequence, the long term benefit of having a healthy partner is now given not only by the equal, but also by the unequal division. 
Since this latter sharing additionally provides a larger amount of immediate resource, being selfish is always the most advantageous strategy.

\section{Discussion}

%
%
The presented results show that generosity can emerge as the optimal strategy of a model for resource gathering and sharing tasks in animals.
Importantly, it emerges even though the agents are interested in maximizing only their personal income, i.e. they are \quotes{greedy}.
However, an equal resource division can be more rewarding than the selfish one in the long run, specifically, when the inequality (\ref{eq:generous_ph}) is satisfied.
The underlying mechanism is that a generous player provides more food to the partner, increasing its health and, thus, making it more efficient at gathering resources in the future, that can be potentially shared with the player.
%
%
It is worth noting that the crucial ingredient that makes the equal resource sharing optimal is the coupling of the sharing task with the resource gathering through the animal health, as mentioned in Section \ref{sec:model}.
To better understand this observation, one can gradually decouple the two tasks by introducing a random component in the health-update function.
For example, with probability $\eta$, the player health is chosen with equal probability to be $0$ or $1$, while, with probability $1-\eta$, the step function used above is employed.
As shown in \ref{sec:generalization}, as $\eta$ increases, the effect that the sharing task has on the resource gathering becomes weaker, and the selfish strategy tends to be the optimal strategy for a larger region of parameters, and becomes the only solution for $\eta=1$.

%
%
The analytical solution of the Bellman optimality equation allows us to identify general rules that govern the fair-selfish dilemma.
First, increasing the discount factor $\gamma$, and therefore putting more weights on the rewards obtained in the future, favours generosity.
This parameters can also be interpreted as the probability that the game repeats at the next step, implying that $(1-\gamma)^{-1}$ is the expected number of episodes to be played, i.e. the time horizon.
Therefore, the longer an agent expects to play, the larger is the parameter space associated to inequity aversion.
The second crucial quantity is the symmetry between the players, described by the specialization parameter $s$.
It can have two interpretations.
The first one is how much more frequently a player performs the task with respect to the other, for example because of dominance ranks within the community.
The alternative interpretation is about the ability in performing the task, in particular the difference in the probability of success of gathering resource between the two players.
It appears that a strong player asymmetry disfavours generosity, possibly leading to the dominance of one player over the other, if (\ref{eq:expl_ph}) is satisfied.
Finally, a crucial role is played by the resource abundance: the generous behaviour is possible only if the environment provides an amount of resources within the window defined by (\ref{eq:abundance_ineq}).
The transition from selfish to generous behavior as a consequence of different resource availability is observed, for example, in the chimpanzees in the Ta\"i National Park. 
They are more generous when the prey is large, typically an adult of a Colobus monkeys, with respect when they capture infants \cite{boesch1989hunting}.
Moreover, it seems that chimpanzees in the rain forest show much more resource sharing than the ones in savanna-woodlands \cite{boesch1989hunting}, and, again, this can be due to a change in the resource availability.

%
%
An important consideration is in order about the simple rule discussed above: generosity is favored by a long time horizon. This closely resembles the condition for cooperation in reciprocity \cite{trivers1971evolution}.
For example, in the classical setting of reciprocity, which identifies Tit for Tat as the cooperative strategy in an iterated prisoner dilemma, cooperation emerges if the probability of interacting again with the same partner (the counterpart of $\gamma$) is sufficiently large \cite{axelrod1981evolution, stephens1996modelling, nowak2006five, whitlock2007costs, buettner2009cooperation}.
This analogy can be traced back to a similar way of thinking about generosity: it emerges from greedy players who want to maximize their personal fitness/return, without considering, for example, the fitness of relatives as in kin selection. Moreover, in both models, generosity can be chosen because it provides an advantage in the long run, while being inefficient at the present time.
%
%
However, there are substantial differences with our approach here.
First, the fundamental mechanism that induces cooperation is different: here it is based on the feedback that the resource sharing has on the next gathering task through the health variable.
In reciprocity the key is to recognize the partner and to remember the game outcome at the previous step, allowing agents to play strategies such as tit-for-tat or win-stay lose-shift \cite{nowak1993strategy}.
We want to stress that our Markov game does not require to remember the previous outcome of the game, but the agent must be able to recognize the health state of his partner, which is more biologically plausible.
Second, here we focus on a specific instance of cooperation: the choice of dividing acquired resources in equal parts, i.e. second order inequity aversion.
To better understand how this  is related with classical theories of cooperation, it is useful to introduce the cost in fitness for the generous action, $c$, and the benefit $b$ that this action provides to the recipient.
If one computes these two quantities for inequity aversion, one finds  $c = R\delta/2$, which is the difference between what a player can potentially acquire by being selfish and what it actually gains by the equal sharing, and $b = R\delta/2$, the recipient receives exactly what the donor looses.
This condition, $c = b$, typically does not allow the emergence of cooperation in evolutionary games.
For example, in the seminal work of Martin Nowak \cite{nowak2006five}, five mechanism for cooperation are introduced and, for each of them, an inequality states when the strategy is evolutionary stable.
In reciprocity, for example, the inequality reads $\gamma > c/b$, where $\gamma$ is the probability that there will be another interaction with the same player.
If $c=b$ cooperation is no longer stable neither for reciprocity nor for all the other four mechanisms.
The last important difference is that, as already stressed above, instead of looking at generosity as an evolutionary outcome, the present work recovers generosity as the best strategy of a decision-making problem that animals can learn through trial and error. 
A crucial advantage of the latter approach is that learning works on much faster time-scales than evolution.
This allows individuals to modify strategies during their life time in response to varying environmental conditions, such as abundances of resources in their habitat or changes in the social ranks within a community.

%
%
The model is clearly oversimplified to address real situations and many additions are conceivable. 
Obviously, the drawback of increasing the model complexity is that the system of equations becomes analytically unsolvable. Nonetheless, it can be approached with numerical methods, such as dynamic programming techniques \cite{sutton1998introduction}.
Just to mention some interesting generalizations, more than two agents can be considered, each with a private policy about the fraction of resource to share with the other players.
Also, the health space can be expanded including intermediate states between the fully healthy and the exhausted player, and more than two choices for the resource sharing can be added.
%
%
In this direction, recent works from the DeepMind lab \cite{leibo2017multi, hughes2018inequity, Wang2018EvolvingIM} study dynamics of cooperation-competition in Markov games inspired by real systems, such as a pair of wolves that has to catch a prey in a grid-world environment, or several agents who have to harvest resources whose spawning rate drops to zero if all the resources have been gathered.
This goes in the direction of adding more realistic details to the game: the system complexity increases, and analytical approaches are no longer possible.
However, there are extremely powerful tools and algorithms to efficiently find rewarding strategies. 
Most of them are based on (deep) reinforcement learning \cite{mnih2015human, li2017deep}.
The problem of generosity and cooperation is just one among many systems in which Markov games, together with reinforcement learning techniques, can find application.
Notably, this framework is also at the basis of several recent successes in \quotes{artificial intelligence}, e.g. \cite{silver2016mastering, moravvcik2017deepstack}.

%
%
The concept of reinforcement learning and its application in Markov decision processes allows us to introduce another important remark.
These kind of algorithms are, generally speaking, grounded on the idea of trial and error.
Importantly, a lot of behavioral and neuroscientific evidence claims that animals can learn using very similar processes, in particular temporal difference algorithms \cite{dayan2008reinforcement, niv2009reinforcement}.
This leads to the biologically reasonable assumption that animals usually learn efficient strategies for the daily-tasks performed in nature.
This can include how to acquire and divide resources, implying that the generous (or the selfish) sharing can be viewed as the optimal policy of a decision-making problem.
Clearly, whether resource sharing and, in general, cooperation between animals are shaped by evolution or by a learning process is an extremely difficult question to answer \cite{lehmann2008social}, with a strong specificity on the species and the behavior under study.
Classical approaches rely mainly on the evolutionary mechanism \cite{nowak2006five}.
Alternatively, our model shows that a subset of these behaviors, i.e. generous resource sharing, can emerge as a result of the learning of a simple Markov game without requiring a population of players and evolutionary time scales.
A reasonable possibility is that what really happens in animals is a combination of the two mechanisms.

%
%
Interestingly, recent successes in obtaining cooperation between AI agents are based on the combination of evolution and learning \cite{Wang2018EvolvingIM}.
The algorithm works on two different time scales.
On the faster time scale, the agent learns with reward function composed of the actual environmental rewards, plus a part of \quotes{intrinsic motivations}.
On the slower evolutionary time scale, a population of agents can evolve the shape of those intrinsic motivations after having completed a learning task, trying to maximize a given fitness function.
As a result, the intrinsic motivations are not connected to real sources of rewards, but greatly help the algorithm to learn efficient strategies \cite{singh2009rewards, singh2010intrinsically}.

%
%
If one assumes that the behavior is mainly driven by learning, then our game can be potentially used to describe and predict dynamics that really happen in animals.
The case of vampire bats would be the most natural one for testing predictions, since their behavior closely resembles our game.
Moreover, it can be reasonable to assume that they learn, because individuals select other specific individuals as sharing partners \cite{wilkinson1990food}, and partner selection can be carried out only within the time scales of learning.
A fascinating possibility would be testing the emergence of different optimal strategies by controlling, for example, the specialization parameter $s$.
To this end, one could force an individual to stay in the roost, therefore introducing asymmetry between the animals in performing the gathering task.
Does this affect the amount of blood transferred during the sharing, leading to some sort of exploitative behavior?
Another possible test could involve the quantity of resource available, and, in particular, if overabundance or scarcity of food would reduce the generous sharing.

As a final remark, we want to mention a further context in which generalizations of our model can be potentially applied.
Let us express in other words the key mechanism for the emergence of generosity:  a selfish action would lead to an exhausted partner, who will get no resource the next time it will play as proposer, making also the first player exhausted and unable to get resource any longer.
Therefore generosity can be seen as an \quotes{insurance} against the community collapse (i.e. the two players exhausted).
This really resembles the mechanism proposed for the emergence of egalitarianism in Bushmen indigenous groups, where there is a strong social pressure towards economic equality and sharing of resources \footnote{We thank an anonymous referee for pointing out this parallelism.}. Anthropologists suggest that this social convention allows the group to survive in adverse environmental conditions \cite{cashdan1980egalitarianism}. 
Differently, other communities which have the possibility to accumulate resources show social stratification and a stronger unbalance of ownership of wealth items. Qualitatively, one can imagine that, in the latter case, inequality (\ref{eq:abundance_ineq}) is violated and selfishness emerges. Clearly, these kind of phenomena require further modelling efforts that go beyond the purpose of the present paper.
However, these qualitative agreements tempt us to speculate that complex features of the primitive human society share a similar backbone mechanism as the one proposed here.

\section*{Acknowledgements}
We are grateful to Agnese Seminara, Jacopo Grilli, Alberto Pezzotta, Matteo Adorisio, Claudio Leone, and Mihir Durve for useful discussions.

\section{Methods}

\subsection{Bellman optimality equation} 
The model can be described as an extension of a Markov Decision Process \cite{howard1960dynamic, sutton1998introduction} for more than one agent: a Markov game \cite{littman1994markov, claus1998dynamics}.
It is defined by a set of states $\mathcal{S}$, and a set of actions that each player can take $\mathcal{A} = \mathcal{A}_1 \otimes \mathcal{A}_2$ (for the two-player case).
From each state $s \in \mathcal{S}$ and set of actions $\vec{a} = (a_1, a_2) \in \mathcal{A}$ the game jumps to a new state $s'$ according to the transition probabilities $p(s'|s,\vec{a}) \in PD(\mathcal{S})$, and each player gets a reward according to the reward function $\vec{r}(s, \vec{a}) \in \mathbb{R}^2$.
The player $i$ chooses which action to take from a given state with a probability given by its policy: $\pi_i(a_i|s) \in PD(\mathcal{A}_i)$.
In Table \ref{tab:mdp} it is shown how the model can be cast into this framework (see \ref{sec:MDP} for a more detailed explanation.)
To compute the optimal strategy, the key quantity to consider  is the quality function $Q_i(s,a_i,\pi_i,\pi_{-i}) = \mathbb{E}_{\pi_i,\pi_{-i}}\left[ \sum_{t=1}^\infty r_i^{(t)} \gamma^{t-1} | S^{(1)}=s, A^{(1)}_i=a_i \right]$, which is the expected return (\ref{eq:return}) of the player $i$ starting from the state $S^{(1)} = s$, choosing the action $A^{(1)}_i = a_i$, by playing  with a policy is $\pi_i$, and by knowing the policy of the other player $\pi_{-i}$.
The optimization problem of maximizing the return (\ref{eq:return}) by knowing that also the other player is optimizing its return simultaneously, can be solved through the following Bellman optimality equation (derived in \ref{sec:solution}):
\begin{equation}
\begin{cases}
\pi^*_i(a_i|s) = \delta (a_i - \underset{b}{\text{argmax}} Q^*_i(s, b) )
\\
Q^*_i(s, a_i) = \mathbb{E}_{p, \pi^*_{-i}} \left[ r_i(s, a_i, a_{-i}) + \gamma\; \underset{b}{\max}Q^*_i(s, b) \right]
\end{cases}
\label{eq:bellman}
\end{equation}
where the optimal policy from the state $s$ is deterministic and consists in choosing the action that maximizes the best quality $Q^*_i(s, a_i)$.

\begin{table}
\centering
\caption{States, actions, rewards and transitions probabilities of the gathering-sharing model.}
\begin{tabular}{cccc}
State, $s$ & Action, $\vec{a}$ & Reward, $\vec{r}$ & Transition probability, $p(s'|s,\vec{a})$ \\
\midrule
$1h$ & $(e,\emptyset)$ & $(\frac{1}{2}, \frac{1}{2})$ & $q \;$ to $1h$, $\;\;$ $1-q \;$ to $2h$ \\
$1h$ & $(u,\emptyset)$ & $\frac{1}{2}(1+\delta,1-\delta)$ & $q \;$ to $1h$, $\;\;$ $1-q \;$ to $2t$ \\
$1t$ & $(e,\emptyset)$ & $(0,0)$ & $q \;$ to $1t$, $\;\;$ $1-q \;$ to $2t$ \\
$1t$ & $(u,\emptyset)$ & $(0,0)$ & $q \;$ to $1t$, $\;\;$ $1-q \;$ to $2t$ \\
$2h$ & $(\emptyset,e)$ & $(\frac{1}{2}, \frac{1}{2})$ & $q \;$ to $1h$, $\;\;$ $1-q \;$ to $2h$ \\
$2h$ & $(\emptyset,u)$ & $\frac{1}{2}(1-\delta,1+\delta)$ & $q \;$ to $1t$, $\;\;$ $1-q \;$ to $2h$ \\
$2t$ & $(\emptyset,e)$ & $(0,0)$ & $q \;$ to $1t$, $\;\;$ $1-q \;$ to $2t$ \\
$2t$ & $(\emptyset,u)$ & $(0,0)$ & $q \;$ to $1t$, $\;\;$ $1-q \;$ to $2t$ \\
\bottomrule
\end{tabular}
\caption{The model is composed of four states identified by the individual playing as proposer, $1$ or $2$, and its health, $h$ for a healthy player, $t$ for an exhausted one.
For each state, the proposer can choose an equal, $e$, or an unequal sharing $u$, while the bystander can only play a fictitious action $\emptyset$.
For each state-action pair the rewards are shown in the third column, and the transition probabilities in the fourth one, where $q = (1+s)/2$ is the probability to choose the first player as proposer.
For more details, see \ref{sec:MDP}}
\label{tab:mdp}
\end{table}

\appendix

\section{Mathematical details of the model}

\subsection{Resource-gathering-sharing model as a Markov game}
\label{sec:MDP}

\begin{figure}[ht]
\includegraphics[width=0.95\textwidth]{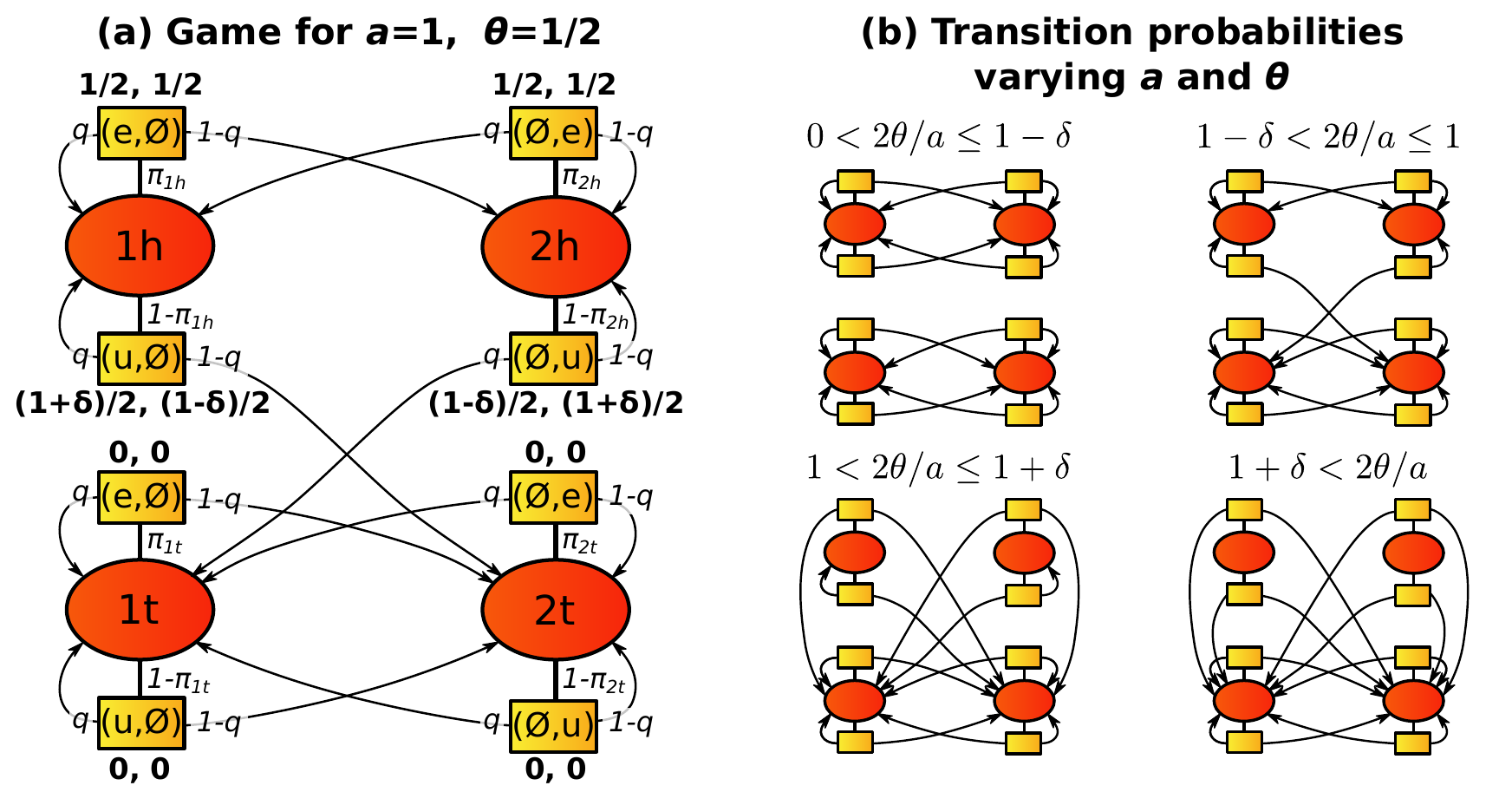}
\centering
\caption{(a): Model as a Markov Decision Process for two players and fixed resource abundance $a=1$ and food threshold $\theta = 1/2$. The four states are the dark orange circles: there are two possible health states, $h$ or $t$, for two possible proposers (the two players), $1$ or $2$.
For each state there are two actions, which are the yellow squares. The label $e$ stands for the equal/generous choice, $u$ for the unequal/selfish one, and $0$ is a fictitious action for the player that is playing as bystander. The pair of bold numbers above or below the action squares are the rewards for the two players as a consequence of that action.
The arrows describe the transitions from one state-action pair to the next state, with probability $q = (1+s)/2$ or $1-q$.
(b): Transition probabilities varying the resource abundance and the food threshold.
}
\label{fig:mdp}
\end{figure}

Let us first describe the model at fixed resource abundance $a=1$ and food threshold $\theta = 1/2$.
The Markov game has four states, each of them characterized by the agent who is playing as proposer and by its health, Figure \ref{fig:mdp}a. 
The following notation will be used: $\mathcal{S} = \lbrace 1h, 2h, 1t, 2t \rbrace$, where the first letter specifies the agent playing as proposer, and the second one its health: $h$ for a healthy individual ($h_i=1$) and $t$ for a tired/exhausted one ($h_i=0$).
From the states $1h$ and $1t$, there are two possible actions, $(e,\emptyset)$ and $(u,\emptyset)$.
The first element is the action of the first agent, who is playing as proposer, and can choose to be equal/generous $e$, or unequal/selfish $u$. 
This choice is determined by its policy $\pi_{1h}$, which is the probability of selecting the generous action from the state $1h$ ($\pi_{1t}$ from $1t$).
The second passive player can only choose a single fictitious action $\emptyset$.
From $2h$ and $2t$ the choice is in the hand of the second player, and therefore the actions are $(\emptyset,e)$ and $(\emptyset,u)$, which are determined by the policies $\pi_{2h}$ or $\pi_{2t}$.
Each action from the states $1t$ and $2t$ (i.e. with an exhausted proposer) provides zero reward/resource-fraction for both the players $\vec{r} = (0,0)$, where the pair indicates the reward for the first and the second player respectively, as indicated in Figure \ref{fig:mdp}a in bold. 
The rewards are instead $(1/2, 1/2)$ for a healthy proposer who plays generously, and $\vec{r}(1h,(u,0)) = 1/2(1+\delta, 1-\delta)$ and $\vec{r}(2h,(0,u)) = 1/2(1-\delta, 1+\delta)$ for the selfish actions of a healthy proposer.
The transition probabilities are determined by the health-update function and the stochastic selection of the proposer.
First, one has to consider that each action leads to a state where the new proposer is the first player, $1h$ or $1t$, with probability $q = (1+s)/2$, and to $2h$ or $2t$ with probability $1-q$ .
Second, the action leads to $1h$ or $1t$ depending on the reward received by the first player: if it is greater or equal than $\theta = 1/2$, then the health-update function leads to a healthy player, and $1h$ is chosen (and similarly the choice between $2h$ or $2t$ depends on the resource fraction received by the second player).
Note that the the generous action of a healthy proposer always leads to a healthy-state ($1h$ or $2h$), since both the player receive an amount of food equal to the minimal food threshold $\theta=1/2$, while the transitions from a selfish-action can lead to an exhausted state ($1t$ or $2t$), if the bystander is chosen to be the new proposer.
Importantly, in this minimal setting, if the game moves to an exhausted-state, it can no longer jump to $1h$ or $2h$.
See Section \ref{sec:generalization} for a more general model, which introduces a probability of health recovery independent of the received resource fraction.

A generic resource abundance $a>0$ and food threshold $\theta>0$ change the reward function and the transition probabilities.
Specifically, all the rewards are multiplied by the parameter $a$, while the transitions show four configurations depending on inequalities involving the ratio $\theta/a$, Figure \ref{fig:mdp}b.
In the first case, $2\theta/a < 1-\delta$, the food threshold is less than the reward of the bystander in a selfish division (performed by a healthy proposer), and, therefore, both the unequal and the equal choice (of healthy proposers) always lead to two healthy players.
In the second case, the food threshold and the resource abundance are set in such a way that the only case when there is exhaustion is being the bystander of unequal division.
Therefore, as a consequence of a selfish action, with probability $1-q$, the game moves in one of the two bottom states (as in the special case $a=1$ and $\theta = 1/2$ discussed above).
When $\theta$ becomes larger that the reward of the generous division, also the equal action leads to poor-health players (third case).
Finally, in the fourth case, when $\theta > a(1+\delta)/2$, the resource abundance in the environment is so scarce that even the resource of a selfish proposer does not allow it to be healthy.

\subsection{Bellman optimality equation for Markov games}

As mentioned in the main text, the aim of each player is to maximize its exponentially discounted return by tuning its policy.
This objective function is defined as:
\begin{equation}
G_i (\pi_i, \pi_{-i}) = \mathbb{E}_{\pi_i, \pi_{-i}} \left[ \sum_{t=0}^\infty \gamma^{t} r_i^{(t+1)}  \right] = 
\sum_{t=0}^\infty \gamma^{t}  \sum_{s \in \mathcal{S}, \vec{a} \in \mathcal{A}} \rho_t(s) \pi_1(a_1|s) \pi_2(a_2|s) r_i(s, \vec{a}) ,
\label{eq:return2}
\end{equation}
where $\rho_t(s)$ is the probability that the game is in the state $s$ at time $t$, and all the other quantities have been introduced in the Method section of the main text.
By taking advantage of the Markov property, one can express the state probability density evolution by considering the following Chapman-Kolmogorov equation:
\begin{equation}
\rho_{t+1}(s') = \sum_{s,\vec{a}} p(s'|s,\vec{a}) \pi_1(a_1|s) \pi_2(a_2|s) \rho_t(s) .
\label{eq:CK}
\end{equation}
Here it is useful to introduce the average residence time in the state $s$: $\eta(s) = \sum_t \gamma^t \rho_t(s)$, and, from now on, to consider Equation (\ref{eq:return2}) dependent on it instead of the state probability density and the summation over the episodes.
The optimization problem here is to maximize the discounted return of each player with respect to its policies.
However, one has to be careful that there is an implicit dependence on them through $\eta(s)$.
This can be approached by considering the average residence time as a new independent variable, maximizing also over it, and imposing a Lagrangian constraint over the following expression, which can be derived by using the Markov process dynamics (\ref{eq:CK}):
\begin{equation}
\eta(s') = \rho_0(s') + \gamma \sum_{s,\vec{a}} p(s'|s,\vec{a}) \pi_1(a_1|s) \pi_2(a_2|s) \eta(s) .
\label{eq:res_time_evol}
\end{equation}
The Lagrangian function to maximize is then:
\begin{equation}
\begin{split}
F_i = & \sum_{s \in \mathcal{S}, \vec{a} \in \mathcal{A}} \eta(s) \pi_1(a_1|s) \pi_2(a_2|s) r_i(s, \vec{a}) - 
\\
& - \sum_{s' \in \mathcal{S}} \left[ \phi_i(s') \left( \eta(s') - \rho_0(s') - \gamma \sum_{s,\vec{a}} p(s'|s,\vec{a}) \pi_1(a_1|s) \pi_2(a_2|s) \eta(s) \right) + \right.
\\
& \left. + \lambda_1(s') \left( \sum_{a_1 \in \mathcal{A}_1} \pi_1(a_1|s') - 1 \right) + 
\lambda_2(s') \left( \sum_{a_2 \in \mathcal{A}_2} \pi_2(a_2|s') - 1 \right) \right] ,
\end{split}
\end{equation}
where $\phi_i(s)$ is the Lagrangian multiplier of the average residence time dynamics (\ref{eq:res_time_evol}), and $\lambda_i(s)$ the multipliers of the policy normalization.
One has to notice that the functional is linear in the policies, implying that the expression does not have a stationary point (the maximum is at the boundary of the policy domain), and imposing the derivative of the functional equal to zero does not lead to any solutions.
To overcome this problem, one can add a regularization term, such as the entropy of the policies: $H_i(s) = - \sum_{a_i} \pi_i(a_i|s) \log \pi_i(a_i|s)$.
The regularized Lagrangian function becomes:
\begin{equation}
F_i^{(\epsilon)} = F_i + \epsilon \sum_s \eta(s) \left( H_1(s) + H_2(s) \right) ,
\label{eq:reg_lagrangian}
\end{equation}
where the parameter $\epsilon$ controls the weight of the entropy with respect the original functional.
At the end of the calculation, one can impose the limit $\epsilon \rightarrow 0$ to recover the non-regularized solution.

The stationary point can be found by setting the derivatives of $F_i^{(\epsilon)}$ with respect to $\pi_i(a_i|s)$ and $\eta(s)$ equal to zero.
This resembles the concept to Nash equilibrium: the solution is the policy that maximizes the return, and from which a change in the strategy is not convenient.
Note that, in general, the system can have more than one solution, as discussed in our specific case later. In that case the best strategy (the Nash equilibrium) is the one that has larger return among the solutions.
The derivative with respect to the policy $i$ leads to the following expression for the best policy:
\begin{equation}
\pi_i^*(a_i|s) = \frac{\exp \left[ Q_i^*(s,a_i)/\epsilon \right]}{\sum_{b_i} \exp \left[ Q_i^*(s,b_i)/\epsilon \right]} ,
\label{eq:best_policy}
\end{equation}
where the Lagrangian constraint $\lambda_i(s)$ has been chosen in such a way that the policy is normalized, and $Q_i^*(s,a_i)$ is defined as:
\begin{equation}
Q_i^*(s,a_i) = \sum_{s', a_{-i}} p(s'|\vec{a},s) \pi^*_{-i}(a_{-i}|s) \left( r_i(\vec{a},s) + \gamma \phi_i(s') \right) .
\label{eq:quality}
\end{equation}
It can be proven that the Lagrangian multiplier $\phi_i(s)$ is the maximal return (\ref{eq:return2}) that the player $i$ can expect starting the game from the state $s$, also called the best \textit{value function}, which, from now on, it is going to be indicated also with $V^*_i(s)$.
As a consequence, the variable $Q^*_i(s,a_i)$ is the best \textit{quality function} of the player $i$, which corresponds to the maximal return of a game starting from the state $s$ and choosing the action $a_i$.
The derivative of the Lagrangian function with respect to the average residence time $\eta(s)$ leads to an expression for the value function:
\begin{equation}
\phi(s) = V_i^*(s) = \epsilon \log \left[ \sum_{a_i} \exp \left[  Q_i^*(s,a_i)/\epsilon \right] \right] + \epsilon H_{-i}(s) .
\label{eq:value}
\end{equation}
Finally, imposing the limit $\epsilon \rightarrow 0$ to the expressions (\ref{eq:best_policy}), (\ref{eq:quality}), (\ref{eq:value}), one recovers the optimality Bellman equations:
\begin{equation}
\begin{cases}
\pi^*_i(a_i|s) = \delta \left( a_i - \underset{b}{\text{argmax}} Q^*_i(s, b) \right)
\\
Q_i^*(s,a_i) = \sum_{s', a_{-i}} p(s'|\vec{a},s) \pi^*_{-i}(a_{-i}|s) \left( r_i(\vec{a},s) + \gamma V^*_i(s) \right) \\
V^*_i(s) = \underset{a_i}{\max} Q^*_i(s,a_i) .
\end{cases}
\label{eq:bellman2}
\end{equation}

\subsection{Exact solution of the model}
\label{sec:solution}

In this section, the optimal policies of the gathering-sharing model are derived through the Bellman equations (\ref{eq:bellman2}).
Since there are four states $\lbrace 1h, 2h, 1t, 2t \rbrace$ and two players, there are eight equations for the value functions.
For example, the ones of the first player read:
\begin{equation}
\begin{cases}
\begin{aligned}
V^*_1(1h) = &  \max \left[   1/2 + \gamma \left( q V^*_1(1h) + (1-q) V^*_1(2h) \right) \right. \; ; 
\\
& \left. (1+\delta)/2 + \gamma \left( q V^*_1(1h) + (1-q) V^*_1(2t) \right) \right]
\end{aligned}
\\
\begin{aligned}
V^*_1(2h) = & \pi_{2h} \left( 1/2 + \gamma \left( q V^*_1(1h) + (1-q) V^*_1(2h) \right) \right) +
\\
& + (1-\pi_{2h}) \left( (1-\delta)/2 + \gamma \left( q V^*_1(1t) + (1-q) V^*_1(2h) \right) \right)
\end{aligned}
\\
V^*_1(1t) = \max \left[ \gamma \left( q V^*_1(1t) + (1-q) V^*_1(2t) \right) \; ; \;  \gamma \left( q V^*_1(1t) + (1-q) V^*_1(2t) \right) \right]
\\
\begin{aligned}
V^*_1(2t) = & \pi_{2t} \gamma \left( q V^*_1(1t) + (1-q) V^*_1(2t) \right) +
\\
& + (1-\pi_{2t}) \gamma \left( q V^*_1(1t) + (1-q) V^*_1(2t) \right) .
\end{aligned}
\end{cases}
\label{eq:values1}
\end{equation}
The equations for the second player can be obtained by taking advantage of the model symmetry: it is invariant by exchanging the two players, $i=1 \rightarrow i=2$, $i=2 \rightarrow i=1$, and the sign of the player specialization $s \rightarrow -s$ (or, equivalently, $q \rightarrow 1-q$). Looking at the last two equations of the system, it is easy to see that the values of the \quotes{tired} states are all zero: $V^*_1(1t) = V^*_1(2t) = 0$, and, similarly, the ones of the second player: $V^*_2(1t) = V^*_2(2t) = 0$.
After some straightforward mathematical manipulations, and adding the expression for the best policy, one can derive the following set of equations for the first player:
\begin{equation}
\begin{cases}
V^*_1(1h) = \frac{1}{2}(1+\delta) + \gamma q V^*_1(1h) + \max \left[ \gamma (1-q) V^*_1(2h) - \frac{\delta}{2} \; ; \; 0 \right]
\\
V^*_1(2h) = \frac{1}{2}(1-\delta) + \gamma (1-q) V^*_1(2h) + \pi_{2h} \left( \frac{\delta}{2} + \gamma q V^*_1(1h) \right)
\\
\pi^*_{1h} = H \left[ \gamma (1-q) V^*_1(2h) - \frac{\delta}{2} \right] ,
\end{cases}
\label{eq:values2}
\end{equation}
where $H[x]$ stands for the Heaviside theta function, which is $1$ for a positive argument and $0$ otherwise.
The system above, together with the other three equations for the second player, defines a closed system of non-linear equations, which can be solved as follows.
One has to define two variables, $x_1 = \gamma (1-q) V^*_1(2h) - \delta/2$, and $x_2 = \gamma q V^*_2(1h) - \delta/2$, which are the arguments of the policies of the two players: $\pi^*_{i} = H[x_i]$.
Therefore, it follows that if $x_i$ is greater than zero, than the optimal strategy of the player $i$ is being generous (or selfish if $x_i < 0$).
Note that this quantity is the difference between two terms.
The first one is the advantage of playing generously, $\gamma (1-q) V^*_1(2h)$: since the action has led to a healthy second player, if it is chosen as proposer at the next step (with probability $1-q$), the first player will get a return equal to $V^*_1(2h)$.
The factor $\gamma$ says that this profit must be discounted because it will be obtained at the next round.
The second term, $\delta/2$, is instead the immediate gain of the selfish action.
Therefore the best strategy can be seen as the balance of these two terms.
By expressing the system (\ref{eq:values2}) as a function of the two new variables, one gets:
\begin{equation}
\begin{cases}
V^*_1(1h) = \frac{1}{1-\gamma q} \left( \frac{1}{2} (1+\delta) + \max [x_1; 0]\right)
\\
x_1 \frac{1 - \gamma(1-q)}{\gamma(1-q)} = \frac{1}{2} - \frac{\delta}{2(1-\gamma q)} + H[x_2] \left( \frac{\delta}{2} + \gamma q V^*_1(1h) \right) , 
\end{cases}
\label{eq:values3}
\end{equation}
and two similar equations for the second player, which, again, can be derived from the system above by using the model symmetry.
At this point, one can consider four different cases depending on the signs of $x_1$ and $x_2$.
Indeed, once $x_1$ or $x_2$ are known to be positive ore negative, the system becomes easily solvable.

\subsubsection{Generous phase}

Let us first consider $x_1 > 0$ or $x_2 > 0$.
This implies that the policies are $\pi^*_{ih} = 1$, which corresponds to a generous strategy for both the players.
By solving the system (\ref{eq:values3}) under this assumption, one finds $x_1 = (\gamma (1 - q) - \delta + \delta\gamma)/(2(1-\gamma))$, and $x_2 = (\gamma q - \delta + \delta \gamma))/(2(1-\gamma))$.
This two expressions must be consistent with the initial choice about the sign of the two variables, leading to the following condition of existence:
\begin{equation}
\gamma > \frac{\delta}{\delta + \min[q, 1-q]} = \frac{2 \delta}{1 - |s| + 2\delta } ,
\label{eq:ce_gen}
\end{equation}
which defines the range of parameters in which the generous solution exists as an optimal strategy.
It is also useful for later considerations to derive the values of this strategy, which, for the first player, read: $V_{1\text{,gen}}^*(1h) = V_{1\text{,gen}}^*(2h) = 1/(2(1-\gamma))$.

\subsubsection{Selfish phase}

In the case of negative $x_1$ and $x_2$ the two players play selfishly.
Under such a condition, the explicit expressions of the two variables are $x_1=(\gamma(1-q)-\delta)/(2(1-\gamma(1-q)))$ and $x_2=(\gamma q-\delta)/(2(1-\gamma q))$, leading to the following condition of existence:
\begin{equation}
\gamma < \frac{\delta}{\max[q,1-q]} = \frac{2 \delta}{1+|s|} ,
\label{eq:ce_self1}
\end{equation}
and the following values:  $V_{1\text{,self}}^*(1h) = (1+\delta) / (2(1 - \gamma q))$, $V_{1\text{,self}}^*(2h) = (1-\delta) / (2(1 - \gamma (1-q)))$.
It can be noted that this condition of existence can overlap with that one of the generous phase, implying the presence of two solutions.
This can be interpreted that, if both the individuals are playing generously, it is not convenient for a player to change the action (its value function will be lower), and, at the same time, it is not convenient for an agent to deviate from the selfish-selfish scenario.
However, one can evaluate which is the best situation by comparing the value functions.
In particular, the generous strategy is the optimal one if $V_{1\text{,gen}}^*(1h) > V_{1\text{,self}}^*(1h)$, and $V_{1\text{,gen}}^*(2h) > V_{1\text{,self}}^*(2h)$.
It is easy to show that this is always the case within the region of parameters in which the two solutions overlap, $\frac{2 \delta}{1 - |s| + 2\delta } < \gamma < \frac{2 \delta}{1+|s|}$.
Therefore, it can be concluded that when the inequality (\ref{eq:ce_gen}) is satisfied, it is always more convenient to play generously, while, the selfish strategy is optimal within the following region:
\begin{equation}
\gamma < \frac{2 \delta}{1 + |s| + 2 \max{(0, \delta - |s|)}} ,
\label{eq:ce_self}
\end{equation}
which is the difference between (\ref{eq:ce_self1}) and (\ref{eq:ce_gen}).

\subsubsection{Exploitation phase}

The final two cases consist in a player being selfish and the other generous.
Here we consider only the case $x_1 < 0$ or $x_2 > 0$, i.e. \quotes{the first player is exploiting the second one}.
The opposite case can be obtained through the model symmetry: the two players are exchanged (and therefore the second player is the selfish one), and the sign of $s$ is switched.
By solving the system, one can found $x_1 = (\gamma(1-q) - \delta (1-\gamma)) / (2(1-\gamma q)(1-\gamma(1-q)))$, and $x_2 = (\gamma q - \delta) / (2(1-\gamma q))$, and the following condition of existence:
\begin{equation}
\frac{2 \delta}{1 + s} < \gamma < \frac{2 \delta}{1 - s + 2 \delta} .
\label{eq:ce_expl}
\end{equation}

\subsection{Generic resource abundance and food threshold}
\label{sec:ab}

Generic values of $a$ and $\theta$ affect the model on two sides: the rewards are all multiplied by the factor $a$, and the transition probabilities change as shown in Figure \ref{fig:mdp}b.
Let us first focus on the top-right configuration of Figure \ref{fig:mdp}b: $1-\delta < 2 \theta / \alpha \le 1$.
In such a region, the transition probabilities are the same of the case discussed above ($a=1$, $\theta = 1/2$), therefore, what changes with respect this special case is just the reward function.
Then, one can take advantage of a useful property of the Bellman equation (\ref{eq:bellman2}): if the solution of a model having rewards $r_i(s,\vec{a})$ are $Q^*_i(s,a_i)$ and $V^*_i(s)$, then, the new solutions of a system where all the rewards are multiplied by the same factor $a$ become $a Q^*_i(s,a_i)$ and $a V^*_i(s)$, and the optimal policies do not change.
As a consequence, the conditions of existence presented before are still valid for all the parameters satisfying $1-\delta < 2 \theta / \alpha \le 1$, while all the value and quality functions are re-scaled by the resource abundance $a$.

To evaluate the optimal policies in the other three cases, one has to compute the quality functions of the individual $i$ when it plays as proposer, $Q^*_i(ih,e)$, and $Q^*_i(ih,u)$, indeed the optimal policy depends on these two quantities, $\pi^*_{ih} = H[Q^*_i(ih,e) - Q^*_i(ih,u)]$.
One finds that, in all the three scenarios, the quality of the selfish choice is always greater, implying the optimality of the unequal division.
For example, when $\theta/a \le 1-\delta$, $Q^*_i(ih,e) = 1/2 + \gamma (q V^*_i(ih) + (1-q) V^*_i((-i)h))$, and, $Q^*_i(ih,u) = Q^*_i(ih,e) + \delta / 2$.

\subsection{Health recovery}
\label{sec:generalization}

\begin{figure}[ht!]
\includegraphics[width=0.9\textwidth]{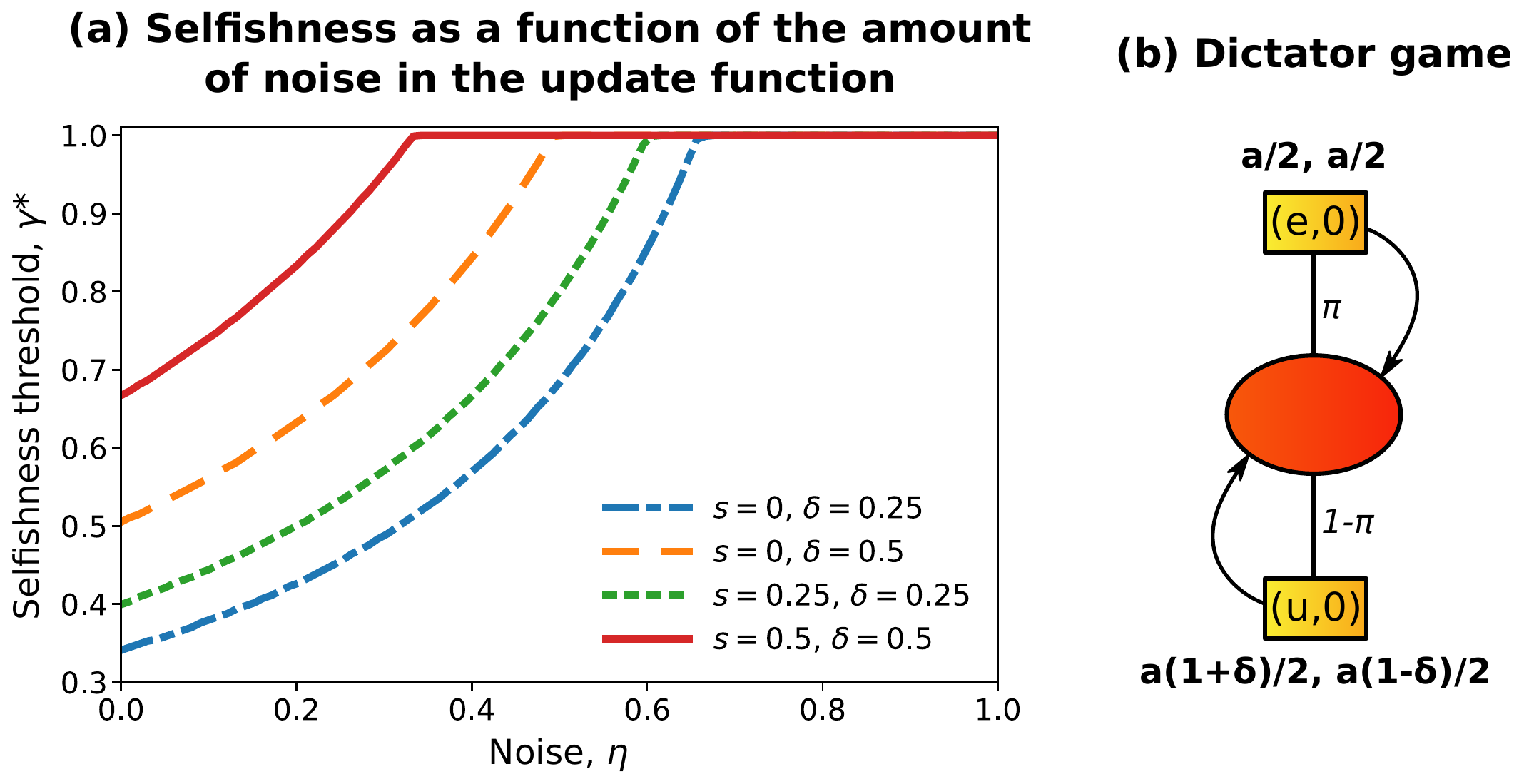}
\centering
\caption{Panel (a): the value of the discount factor $\gamma$ at the boundary of the generosity phase, i.e. the generosity threshold $\gamma^*$, is plotted as a function of the probability that the health is randomly updated, $\eta$, for four choices of the other two parameters $s$ and $\delta$.
Above the lines, the optimal strategy is being generous, while below it can be the selfish or the exploitation one.
The four lines are interpolations of numerically found values of $\gamma^*$, varying the \quotes{noise} $\eta$, and fixing the specialization $s$, and the inequity strength $\delta$.
Specifically, the Markov game is solved with a dynamic-programming algorithm.
Panel (b): classical dictator game with two actions, which can be seen as a particular case of the generalized gathering-sharing model with $s=1$, $\eta = 1$, $\rho = 1$.
One fixed proposer has a certain \quotes{capital} of resource $a$ which has to be divided with a passive bystander.
The two actions are the usual ones: being generous and splitting the capital in equal parts, or selfish and keeping a larger fraction.
After the division the game can repeat with the same rules (neither the proposer change nor the amount of resource). }
\label{fig:gen_vs_noise}
\end{figure}

This section introduces a possible generalization of the game, based in the new following health-update function:
\begin{equation}
h(r) = 
\begin{cases}
H[r-\frac{1}{2}] & \text{with probability} \;\; 1-\eta
\\
\text{Bern}[\rho] & \text{with probability} \;\; \eta ,
\end{cases}
\label{eq:h_update_gen}
\end{equation}
where, with probability $1-\eta$, the health update is the step function $H[r-1/2]$, while, with probability $\eta$, the health is instead randomly chosen according to a Bernoulli variable which is $1$ with probability $\rho$ or $0$ otherwise.
For $\eta = 0$ the previous model is recovered (for simplicity we consider $\theta = 1/2$).
The opposite case, $\eta = 1$, defines a game in which there is no correlation between the amount of food obtained from the sharing task and how the health will be updated, as if the type of resource does not affect the animal health.
This absence of correlation, in turn, implies that the amount of resource gathered at the next episode does not depend on the outcome of sharing task.
In the main text, it is claimed that the link between sharing and gathering is a crucial ingredient to have generosity, therefore, one can expect that uncoupling the two tasks by increasing $\eta$, inequity aversion tends to disappears.
This is shown in Figure \ref{fig:gen_vs_noise}a: the value of discount factor above which the optimal solution is being generous, which we call \textit{selfishness threshold} $\gamma^*$, is numerically computed and plotted as a function of the \quotes{noise} $\eta$.
It has been tested that those lines are independent of the value of $\rho$.
The panel shows that increasing the probability that the health is randomly updated $\eta$, the parameter-space volume in which the optimal strategy is generosity tends to decrease (i.e. $\gamma^*$ increases), eventually reaching $\gamma^* = 1$, which implies that choosing an equal resource division is always inefficient.
Consistently, it can be easily proven that the completely-random case, $\eta=1$, always leads to a selfish scenario.
Indeed, the quality functions of the player playing as proposer $i$ are: $Q^*_i(ih,u) = \delta/2 + Q^*_i(ih,e)$ for any parameter choice, implying that $\pi^*_{ih} = 0$.
It is worth mentioning that the random case with $q=1$ (the proposer is always the same player) and $\rho = 1$ (there is only one health state and, at each step the quantity of resource to divide is fixed to $a$) can be described as composed of only one state.
This recovers the classical \textit{dictator game}, in which one fixed player has to choose how to share a fixed amount of resource with a second passive player, Figure \ref{fig:gen_vs_noise}b.
In this case one clearly has always the selfish strategy as the optimal solution.

\subsection{Modelling the resource sharing with the ultimatum game}
\label{sec:UG}

\begin{figure}[ht!]
\includegraphics[width=0.45\textwidth]{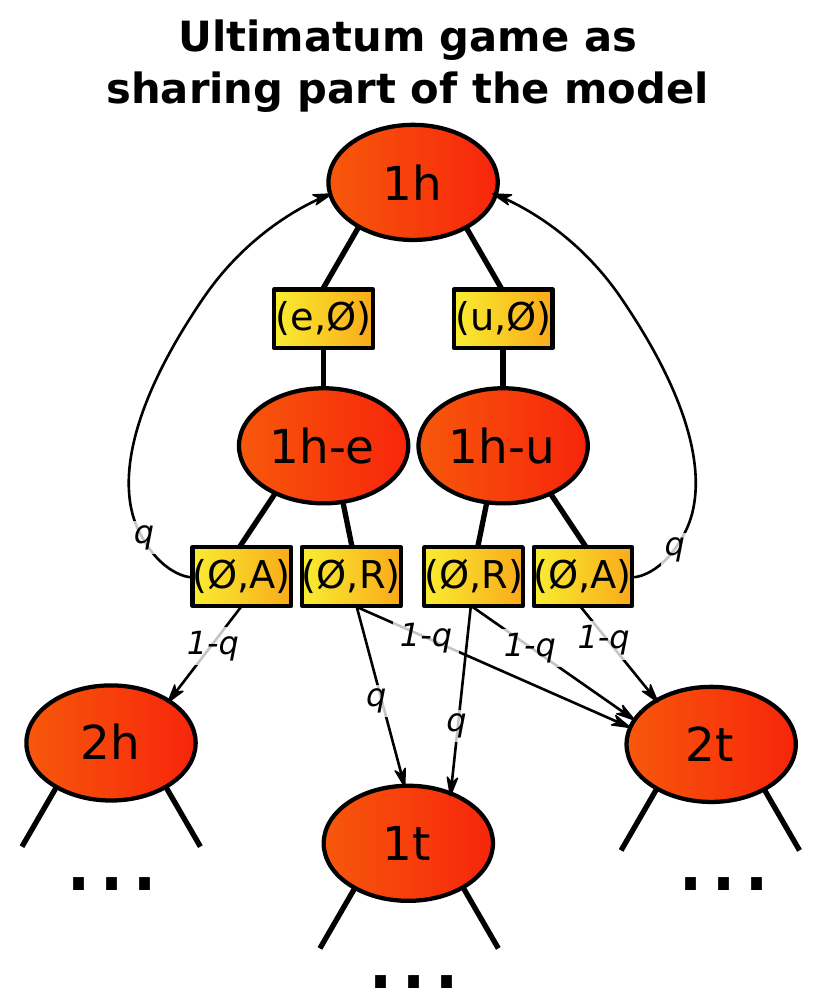}
\centering
\caption{Markov decision process for the gathering-sharing game having the ultimatum game as the sharing module. The total number of states is $12$, but only $6$ of them are drawn.
Specifically, the three corresponding to the ultimatum game from the configuration in which the first player is a healthy proposer, $1h$, and the initial states of the other three configuration $2h, 1t, 2t$.}
\label{fig:UG}
\end{figure}

Here we consider a variant of the model having an ultimatum game to describe the resource sharing.
The ultimatum game can be characterized by three states, which are $1h$, $1h$-$e$, $1h$-$u$ of figure \ref{fig:UG}, which shows the sharing task from the state in which the first player is a healthy proposer.
In the first state, $1h$, the proposer chooses between the equal, $e$, or the unequal, $u$, action in order to divide an amount of resource (the bystander does not take actions at this stage).
Each action makes the game moves to a different next state, one for the equal, $1h$-$e$ and one for the unequal action, $1h$-$u$. 
Those transitions provide zero reward and are deterministic.
From these two states the bystander has the choice to either accept, $A$ or reject, $R$.
Note that having two different states corresponding to the two proposer choices, the bystander can base its decision as a consequence of the received proposal.
If it rejects both the players do not get any resource (from both the two states).
Note that, as a consequence, the health of both the players drops to zero and the game moves to one of the two states having a tired proposer (and zero value): $1t$, or $2t$.
If the bystander accepts the received reward depends on the proposer choice: $r_p = r_b = a/2$ from the state corresponding to the equal choice $1h$-$e$, or $r_p = a/2 (1+\delta)$ and $r_b = a/2 (1-\delta)$ from the state of the unequal action, $1h$-$u$.
The transitions to the next states can be computed as in the base model using the health update function dependent on the received resource, and the probability that the first player is the proposer $q = (1+s)/2$.

By computing the quality functions corresponding to the state-action pairs, one can immediately note that the Nash equilibrium of this model is the same of the one having the simpler dictator game for the resource sharing\footnote{To really establish the equivalence one has to notice that the ultimatum game involves two temporal steps instead of the single step of the dictator game. Therefore, the reward obtained at the end of the game is discounted by a factor $\gamma^2$. One possibility to establish is not to discount the first transition (discount factor equal to $1$), which implies that the value, for example, of $1h$-$e$ is exactly equal to the quality of the state $1h$ and the action $e$.}.
This is due to the fact that the only reasonable choice for the bystander is always to accept in all the cases.
For example, the qualities of the second player from the state $1h,u$ (where the first player is the proposer who has chosen the unequal sharing) are:
\begin{equation*}
\begin{aligned}
Q^*_2(A|1h,u) = & \frac{a}{2}(1-\delta) + \gamma (q V^*_2(1h) + (1-q) V^*_2(2t)) = 
\\
= &\frac{a}{2}(1-\delta) + \gamma q V^*_2(1h)
\\
Q^*_2(R|1h,u) = & \gamma (q V^*_2(1t) + (1-q) V^*_2(2t)) = 0 , 
\end{aligned}
\end{equation*}
where we used the fact that the values of the \quotes{tired} states are zero as previously shown.
This implies that $Q^*_2(A|1h,u) \ge Q^*_2(R|1h,u)$ (where the equality holds only for uninteresting extreme cases), and therefore the optimal strategy is always to accept.
In words, the reject action is always inconvenient since it would make both the proposer and the bystander exhausted and does not provide any immediate reward.
If the policy of the bystander is set always to \quotes{accept}, the game becomes equivalent to the dictator game, proving the equivalence between this and the model discussed so far.

\bibliographystyle{unsrt}  
\bibliography{bib}

\end{document}